\documentclass[aps,twocolumn]{revtex4-2}

\usepackage{hyperref}
\usepackage{graphics,graphicx,epsfig}
\usepackage{amssymb,color}
\usepackage{epsf,epstopdf,wrapfig}
\usepackage{mciteplus}
\usepackage{graphicx,color}
\usepackage{amsmath,amssymb}
\usepackage{dcolumn}   
\usepackage{bm}        
\usepackage{multirow}
\usepackage{dsfont}
\usepackage{lipsum,afterpage}
\usepackage{ulem}

\def\be{\begin{equation}}
\def\ee{\end{equation}}
\def\bea{\begin{eqnarray}}
\def\eea{\end{eqnarray}}

\newcommand{\aref}[1]{\hyperref[#1]{Appendix~\ref{#1}}}

\newcommand{\Z}{\mathbb{Z}}

\newcommand{\One}{1\mkern -4.5mu \rm{I}}

\newcommand{\ke}[1]{|    #1    \rangle}

\DeclareMathOperator\erfc{erfc}

\begin{document}

\title{TASEP Exit Times}

\author{Jérôme Dorignac}
\author{Fred Geniet}
\author{Estelle Pitard}
\affiliation{ Laboratoire Charles Coulomb (L2C), Universit\'e de Montpellier, CNRS, 34095 Montpellier, France \\}
\date{\today}

\begin{abstract}
We address the question of the time needed by $N$ particles, initially located on the first sites of a finite 1D lattice of size $L$, to exit that lattice when  they move according to a TASEP transport model. Using analytical calculations and numerical simulations, we show that when $N \ll L$, the mean exit time of the particles is asymptotically given by $T_N(L) \sim L+\beta_N \sqrt{L}$ for large lattices. Building upon exact results obtained for 2 particles, we devise an approximate continuous space and time description of the random motion of the particles that provides an analytical recursive relation for the coefficients $\beta_N$. The results are shown to be in very good agreement with numerical results. This approach sheds some light on the exit dynamics of $N$ particles in the regime where $N$ is finite while the lattice size $L\rightarrow \infty$. This complements previous asymptotic results obtained by Johansson in \cite{Johansson2000} in the limit where both $N$ and $L$ tend to infinity while keeping the particle density $N/L$ finite. 

\end{abstract}

\maketitle

\section{Introduction}

The TASEP model (Totally Asymmetric Simple Exclusion Process) is a unidirectional model of transport of particles with exclusion on a one dimensional lattice \cite{MacDonald1968}.
It has various interesting applications in traffic on lanes, waiting times lists, directed transport of particles through channels and more \cite{Schadschneider:1564539, Arita2009, ARITA2014}. It can also be mapped on models of interface growth \cite{Johansson2000,Sasamoto_2007}, providing alternate interpretations of its results. 
Originally introduced in the context of the kinetics of biopolymerization, it has also been a paradigmatic model in the field of biological transport since \cite{CHOWDHURY2000199,Zia2011}.

Most theoretical investigations of the TASEP model have been dedicated to obtaining results at stationarity when the flux of particles entering and exiting the lattice has reached a stationary value. In that respect, particle density and current properties have been thoroughly studied \cite{Derrida1992,Derrida1993DE,Lazarescu2011,DERRIDA199865, Chou2011}. 
But some results have also been obtained in non-stationary regimes, especially in infinite lattices. For instance, the exact Green functions of the continuous time TASEP model on $\Z$ have been obtained by Schütz in \cite{Schutz1997}. Related quantities have subsequently been used to determine some asymptotic features of the time evolution of the particle density when starting from a step-initial condition where particles initially populate the left half of the lattice only \cite{Rakos2005}. Much in the same vein, the statistical features of the motion of certain (tagged) particles along the lattice have been elucidated as well \cite{ Sasamoto_2007,NAGAO2004487}. The question we address here pertains to that class of non-stationary problems: how a set of particles, transported according to the TASEP rules, evacuate a finite lattice, especially when they start from a "step-like" configuration where all of them are located on the leftmost sites of that lattice? To answer that question, we shall study the distribution of their exit time and, more specifically, their mean exit time.

Studies on exit times (also called evacuation times or escape times) in single-file systems, that is in 1D systems where particles cannot pass each other, generally involve bidirectional motion like in single-file diffusion (SFD) problems (see for instance \cite{Grabsch2023, PhysRevLett.113.078101} and references therein). In this context, exit time distributions may be analyzed via the first passage time density of a "tracer" (or tagged) particle moving within a crowd of like particles (see e.g. \cite{Sanders2012}). Analysis of these SFD problems shows that the tracer position $x(t)$ has a subdiffusive behaviour leading to a mean-squared displacement that scales as $\langle (x(t)-x_0)^2 \rangle \propto t^{2H}$ at long times where $H$ is the Hurst exponent \cite{Sliusarenko2010,Molchan1999}. This behaviour, due to crowding effects generated by 1D confinement at a given density of particles, contrasts with the $\langle (x(t)-x_0)^2 \rangle \propto t$ scaling typical of a diffusive behaviour for which $H=1/2$. In the particular case of the symmetric exclusion process (SEP) for instance - a 1D hard-core lattice gas problem equivalent to the TASEP model but where particles may equally jump to the right or to the left provided the corresponding site is empty (see e.g. \cite{Derrida2009SSEP}) - the Hurst exponent is $H=1/4$ and the SEP problem has been shown to be equivalent to a fractional Brownian motion (fBm) that depends on the particle density \cite{Gonçalves2008}.

The exit of particles following TASEP transport rules from a finite size lattice share some similarities with SFD systems. In particular, the motion of a given particle is hindered by others (exclusion) and therefore cannot perform a simple independent random walk. But there are    
two main differences between SFD problems and the question we address in this paper. First, the particle density does not remain constant over time because particles progressively leave the system they start from and free the motion of those that remain within the lattice. In that respect, the escape of colloidal particles from microfluidic channels studied in ref. \cite{Locatelli2016} is a problem closer to ours. Second, the motion of particles is unidirectional (particles only move to the right) that is, transport is totally biased. This situation is similar to emergency evacuation in trains or aircrafts where individuals have to quickly walk down a narrow seat aisle \cite{HUANG2018,CAPOTE2012}. The evacuation of particles according to TASEP rules might therefore provide some insight on emergency evacuation although pedestrian dynamics has a quite complex stochastic structure \cite{Jelic2012,KIRCHNER2002260}.

In this paper, we focus on a special setting of the TASEP model: particles start from a step-like initial state and
no particle is injected at site 1. Moreover, all particles exit the lattice as they reach site $L+1$ (absorbing condition after the $L$th site). At time $t=0$, the $N$  particles are located on sites $1,..N$ with $N \le L$, as displayed in Figure \ref{fig:initial_configs}.
We are interested in the emptying time of this model which is equal to the exit time of the leftmost particle of the lattice. In particular, we shall use analytical calculations and numerical simulations to determine the mean exit time (MET).

After introducing the model and quantities of interest in Section \ref{sec2}, we present some exact results for 1 and 2 particles in Section \ref{sec:1particule} and Section \ref{sec:2particule2}. We then use a continuous space and time description of the relative motion of the particles with respect to the leading one to calculate the exit time in the large $L$ limit in Section \ref{sec5}. This approach provides a simple physical approximate solution of the problem, yielding a recursive expression for $T_N(L)$ in the large $L$ limit. These results are then compared to Gillespie simulations  in Section \ref{Gillespie}. In section \ref{conclusion}, we discuss our asymptotic results and compare them to those of Johansson \cite{Johansson2000} obtained in the finite density regime.

\section{Transport model and its exit time distribution} \label{sec2}
\subsection{The TASEP Model}

\begin{center}
\begin{figure}[h]
\includegraphics[scale = 0.4]{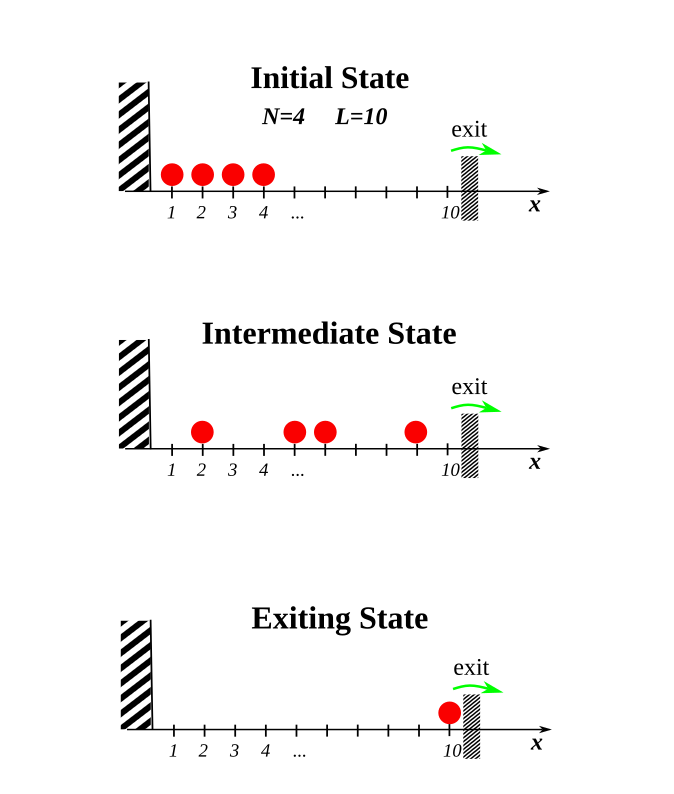}
\caption{Initial, intermediate and final configurations. }
\label{fig:initial_configs}
\end{figure}
\end{center}

The Totally Asymmetric Simple Exclusion Process (TASEP) is a paradigmatic dynamical model for the unidirectional transport of particles on a lattice that takes into account exclusion. It is generally  defined by the following rules: a particle may be loaded on the lattice with probability rate $\alpha$ provided 
the first site is empty. It then proceeds forward with a hopping rate $p$ provided the neighbouring site (on the right) is empty and leaves the last lattice site with a probability exit rate $\beta$. In what follows, we shall study the exit time distribution of $N$ particles initially located on the first (leftmost) $N$ sites of a lattice containing $L\geq N$ sites, see figure \ref{fig:initial_configs} for a pictorial view. We shall moreover assume that the hopping rates on the lattice are homogeneous, $p=\beta$ and using $1/p$ as unit of time, we shall simply set $p=\beta=1$. Finally, the incoming rate $\alpha$ is set to zero in such a way that no particle enters the lattice from $t=0$ onward.

The TASEP model is a Markov process governed by the master equation,
\begin{equation} \label{METASEP}
\frac{d\ke{P(t)}}{dt} = M \ke{P(t)}
\end{equation}
where the probability vector may be written as
\begin{equation} \label{Probavect}
\ke{P(t)} = \sum_{{\bm \sigma}} P_{{\bm \sigma}}(t)\ke{{\bm \sigma}}\,.
\end{equation}
Here, the configuration vector is $\ke{{\bm \sigma}}=\ke{\sigma_1}\otimes\dots\otimes\ke{\sigma_L} $ with column vectors $\ke{\sigma_i}=(1-\sigma_i,\sigma_i)^T$ where $\sigma_i=1$ when site $i$ is occupied by a particle and $\sigma_i=0$ otherwise. The sum runs over all $2^L$ particle configurations. Within our settings, the Markov matrix $M$ reads \cite{RAGOUCY2017}
\begin{equation} \label{MatrixMdef}
M = \sum_{i=1}^{L-1} \One^{i-1}\otimes m \otimes 
\One^{L-1-i} + \One^{L-1} \otimes b\, .
\end{equation}
where $\One$ is the $2\times 2$ identity matrix and where
\begin{equation} \label{Matrixmandb}
m = \begin{pmatrix}
0 & 0 & 0 & 0 \\
0 & 0 & 1 & 0 \\
0 & 0 & -1 & 0 \\
0 & 0 & 0 & 0 
\end{pmatrix};\
b = \begin{pmatrix}
0 & 1  \\
0 & -1 
\end{pmatrix}.
\end{equation}
It is worth noting that, as the incoming rate $\alpha$
has been set to zero, $M$ is a $2^L\times 2^L$ upper triangular matrix. 

\subsection{Exit time distribution} \label{ETDsection}
As the TASEP model is a random process, the time $t$ taken by $N$ particles to empty an $L$-site lattice is a random variable. We shall denote by $p_{N,L}(t)$ its probability density function (PDF). In the terminology of the previous section, the probability that the lattice is empty at time $t$ is given by $P_{\bm 0}(t)$ where ${\bm 0}=(0,\dots,0)$ is the configuration where the $L$ sites are empty. Now, the lattice is empty at time $t$ if the $N$ particles have evacuated it by a time $\tau\leq t$. Then, the probability ${\rm Pr}(\tau \leq t)$ that the exit time of the $N$ particles is less than $t$ is exactly equal to $P_{\bm 0}(t)$. The exit time PDF, $p_{N,L}(t) = d{\rm Pr}(\tau \leq t)/dt$, is therefore given by
\begin{equation} \label{pext}
p_{N,L}(t) = \dot{P}_{\bm 0}(t) = P_{(0,\dots,0,1)}(t) \, ,
\end{equation}
where the dot denotes the time derivative and where the last equality is obtained from the master equation \eqref{METASEP}. It is thus sufficient to evaluate the probability that only site $L$ is occupied at time $t$ to obtain the exit time PDF. 

Taking the Laplace transform of the master equation \eqref{METASEP}, we obtain the following algebraic system
\begin{equation} \label{LTMETASEP}
(s - M) \ke{\tilde{P}(s)} = \ke{P(0)}\, ,
\end{equation}
where $s$ is the parameter of the Laplace transform defined by 
\begin{equation} \label{LTf}
\tilde{f}(s) = \int_0^{\infty}\! dt\, f(t) e^{-st}\, .
\end{equation}
In Eq. \eqref{LTMETASEP}, $\ke{P(0)}$ is the initial probability vector with a single nonzero component:  $P_{(1,\dots,1,0,\dots,0)}=1$ with $N$ 1's and $(L-N)$ 0's.
Solving the triangular algebraic system \eqref{LTMETASEP}, one obtains the Laplace transform of $P_{(0,\dots,0,1)}(t)$ and, from there, the PDF $p_{N,L}(t)$ itself. Of particular interest is the Mean Exit Time (MET) of that distribution. In the next sections, we shall be mainly interested in the asymptotic behavior of this quantity as $L$ becomes large while $N$ remains finite. For that reason, we shall denote the mean exit time of $N$ particles from a lattice  $\{1,L\}$ with $L$ sites as $T^{(f)}_N(L)$.
The superscript $(f)$ emphasizes the fact that this MET is obtained for a finite lattice with $L$ sites and not for a section with $L$ sites
$[1,L]$ embedded in an infinite lattice. We shall come back to that point in section \ref{sec:infinitelattice}. Let us just note for now that $T^{(f)}_N(L)$ may be directly derived from the Laplace transform of $p_{N,L}(t)$ as
\begin{equation} \label{MET}
T^{(f)}_N(L) = - \left. \frac{d\tilde{p}_{N,L}(s)}{ds}\right|_{s=0}\, .
\end{equation}

A word is in order here. The solution of \eqref{LTMETASEP} is technically immediate, both because the system is triangular and because the results are rational fractions in $s$ whose inverse Laplace transforms are straightforward. Nonetheless, it allows for an analytical determination of the exit time distribution $p_{N,L}(t)$ and its MET for small lattices only. The size of the Markov matrix grows indeed exponentially fast with $L$ and we have not found any compact way to express the analytical solution in the general case ($N\leq L$). Of course, starting from $N$ particles, only configurations with at most $N$ particles contribute to the dynamics of the system. The dimension of the Markov matrix reduced to these configurations is much smaller than $2^L$: for instance for $N=2$, the total number of configurations with at most two particles is $L(L+1)/2+1$ which grows algebraically as $L^2/2$ for large $L$. For $L=20$, the reduced Markov matrix is then roughly 200$\times$200 vs $10^6\times 10^6$ for the full one. However, in spite of this drastic reduction, analytical expressions become very lengthy whenever $L>20$ and, although exact, they are not particularly helpful in determining asymptotic behaviors for large $L$. They provide results that can be used as benchmarks for simulations though. Examples of such results for $N=2,3$ are provided in appendix \ref{ERSL}. In the next section, we shall therefore use a different method to tackle the determination of the MET for arbitrary large lattices.

\section{One particle: ballistic regime} \label{sec:1particule}
We briefly treat here the exit time distribution of a single particle initially located on site 1 of an $L$-site lattice $\{1,L\}$. As it is more convenient, we switch from the "Eulerian" description based on particle configurations, that we have used so far to express probabilities, to a "Lagrangian" approach where particles are traced. Let us then call $P(n;t)$ the probability that the particle lies on site $n \in \{1,\dots,L+1\}$ at time $t$. The addition of a virtual $(L+1)$th site allows the particle to exit the lattice. This site is "absorbing" and $P(L+1;t)$ is then the probability that the lattice $\{1,L\}$ is empty.  According to Eq. \eqref{pext}, we then have $p_{1,L}(t)=P(L;t)=\dot{P}(L+1;t)$. 
In the Lagrangian terminology, the master equation \eqref{METASEP} translates into 
\begin{eqnarray} \label{ME1particle}
\dot{P}(1;t) &=& - P(1;t) \\
    \dot{P}(n;t) &=& P(n-1;t) - P(n;t)\, ,\ n \in \{2,\dots,L\}\, .
\end{eqnarray}
Taking the Laplace transform of equation \eqref{ME1particle} with an initial condition given by $P(1;0)=1$ (all other component being zero) yields $\tilde{P}(n;s)=(1+s)^{-n}$. Hence, 
\begin{equation} \label{pexs1part}
\tilde{p}_{1,L}(s) = \tilde{P}(L;s) = (1+s)^{-L}\, ,
\end{equation}
which upon inversion yields the exit time distribution of a single particle out of the $\{1,L\}$ lattice,
\begin{equation} \label{pext1part}
p_{1,L}(t) = \frac{t^{L-1}}{(L-1)!}e^{-t}\, .
\end{equation}
This distribution is of the Poisson type, as expected: a single particle indeed never experiences exclusion and spends on each site a time that  follows the same exponential distribution ($e^{-t}$). Therefore, the total amount of time it spends on the lattice $\{1,L\}$ is nothing but the sum of $L$ exponentially distributed variables which leads to the Poisson distribution \eqref{pext1part}. Moreover, according to equations \eqref{MET} and \eqref{pexs1part}, the mean exit time of the particle is 
\begin{equation} \label{MET1part}
T^{(f)}_1(L) = L \, .
\end{equation}
The particle spends on average a unit of time on each site and thus travels at constant velocity. In that respect, its motion is ballistic. The purpose of the next section is to detail how this motion is hindered when another particle seats initially next to its right side.

\section{Two particles : exact and asymptotic expressions for the MET. }
\label{sec:2particule2}

\subsection{Finite lattice} \label{sec:finitelattice}

Let us label the particles in their exiting order, namely from right to left, and consider first the same problem as in the previous section but with 2 particles initially located on site 2 (first particle) and on site 1 (second particle) of the lattice $\{1,L\}$. We can show (see appendix \ref{MET2partfinite}) that the mean exit time of these two particles is exactly given by 
\begin{equation} \label{MET2partf}
  T^{(f)}_2(L)= L + \frac{L-1}{4^{L-2} } \times \binom{2L-3}{L-1} .
\end{equation}
where $\binom{m}{n}=m!/(n!(m-n)!)$ is the binomial coefficient.
Asymptotically, for large $L$, we then find 
\begin{equation} \label{eq:TF2L}
T^{(f)}_2(L) =  L + \frac{2}{\sqrt{\pi}}\, \sqrt{L} +{\cal O}(L^{-1/2})\,       .
\end{equation}
Comparing this expression to the 1 particle MET \eqref{MET1part} shows that the main effect of adding a particle next to the first one at the start of the process is to delay its exit by an amount of time that is proportional to the square root of the distance it has to travel to exit. In the next section, we shall interpret that result as a consequence of the random motion of the second particle confined on its right side by the random motion of the first one that it cannot overtake.

\subsection{Infinite lattice}
\label{sec:infinitelattice}

We now consider a problem related to the previous one although slightly different: what is the time $T_2(L)$ necessary for 2 particles to exit the section $[1,L]$ of an \textit{infinite lattice}, with the same initial positions as for the finite lattice $\{1,L\}$? This problem has much simpler boundary conditions than in \ref{sec:finitelattice} as particles keep moving on the infinite lattice instead of being absorbed at site $(L+1)$. This will enable us to develop a connection with a diffusion equation. The physical difference between the two situations is that in this present case, a particle having exited the $[1,L]$ section still hinders the previous ones, whereas in the finite domain problem, the dynamics of a particle changes to a ballistic one each time its predecessor exits the lattice $\{1,L\}$. However in the large $L$ limit, we expect the particles mean relative distances to become large, and the additional constraint provided by the following particles to be weak. Our following results will sustain this claim. 

We first use results developed in \cite{Schutz1997}, which provides  an exact formula for the probability of 2 particles to be at positions $X_1$ and $X_2$ at times $t$ knowing their initial positions $Y_1$ and $Y_2$ at time 0. From that we are able to deduce (see appendix \ref{MET2partinfinite}):
\begin{equation}
   T_2(L) =  L + \frac{2}{\sqrt{\pi}} \times \frac{\Gamma(L+1/2)}{\Gamma(L)}
\end{equation}
and this again yields the same asymptotic behaviour than Equation (\ref{eq:TF2L}) :
\begin{equation}
T_2(L) =  L + \frac{2}{\sqrt{\pi}}\, \sqrt{L} +{\cal O}(L^{-1/2})
\end{equation}
therefrom showing the equivalence of the finite and infinite formulation of the exit problem in the large $L$ limit.

We now make a connection between this problem and a diffusion equation. The master equation for this two particles case writes \cite{Schutz1997}:
\begin{eqnarray}
 & & \partial_t P(k_2,k_1;t)  = \nonumber \\ & &  + P(k_2-1,k_1;t)  +  P(k_2,k_1-1;t)  -2  P(k_2,k_1;t)\nonumber \\
 & & P(k,k+1;t)=P(k,k;t) \nonumber \\
 & & P(1,2;t=0)= 1 , \qquad {\rm or \; else \;  0 }
\end{eqnarray}
valid for any $k_2<k_1$, the positions of the trailing and leading particles, respectively.  The boundary condition elegantly accounts for the special case $k_1=k_2+1$. The probability ${\cal P}(\chi;t)$ that the distance between the 2 particles be $\chi$ at time $t$ then follows from
\begin{equation}
    {\cal P}(\chi;t)=\sum_{k_2\geq 1} P(k_2, k_2 + \chi;t)
\end{equation}
and satisfy 
\begin{equation}
  \partial_t {\cal P}(\chi;t)= {\cal P}(\chi+1;t)+{\cal P}(\chi-1;t) -2 {\cal P}(\chi;t) \label{eq:Masterrelativepart}  
\end{equation}
with the boundary condition 
\begin{equation}
{\cal J}(\chi=0;t) \equiv {\cal P}(0;t)-{\cal P}(1;t) = 0 \label{eq:discrete_noflux}
\end{equation}
We notice that this is a discretized version of the diffusion equation with a no flux condition (${\cal J}=0$) originating from the exclusion constraint, and this will enable us to develop an approach based on this equation in the next section.

Solving equation (\ref{eq:Masterrelativepart}) we obtain for the mean distance (see appendix \ref{MRD2partinfinite})
\begin{equation} \label{MeanDistance2part}
    \langle \chi(t) \rangle = \frac{e^{-2t}}{2}[(4t+1)I_0(2t)+4tI_1(2t)] +\frac{1}{2}
\end{equation}
where $I_k$ is the modified Bessel function of order $k$. Consequently, for large times $t$, the distance between the 2 particles varies like
$\langle \chi(t) \rangle \simeq \frac{2}{\sqrt{\pi}}\sqrt{t}$.
In the context of the exit of 2 particles initially at  $k_2=1$ and $k_1=2$, particle 1 reaches the end of the lattice after a time $L$, time at which particle 2 is on average at a distance $\langle \chi(L) \rangle \propto L^{1/2}$ behind particle 1. 
Then particle 2 reaches the end of the lattice with a delay $\langle \chi (L) \rangle$. Finally, for $L$ large, the exit time of the 2 particles is $ T_2(L) \simeq  L + 2\sqrt{L/\pi}$, which is consistent with our previous exact results.

\section{The diffusion approximation}
\label{sec5}
The former calculations suggest the following simple physical approach : since the leading particle has on average a ballistic motion with a constant velocity, it is convenient to study the motion of the rear particles in the reference frame of the leading one. As we saw in equation (\ref{eq:Masterrelativepart}) this leads to a diffusion equation for the motion of the second particle with a no flux boundary condition accounting for the exclusion. This can be  generalized to any of the $(N-1)$ trailing particles, the preceding particle acting as an impenetrable wall due to exclusion, to recursively find the average position of the n-th particle with respect to the leading one.
\begin{center}
\begin{figure}[h]
\includegraphics[scale = 0.5]{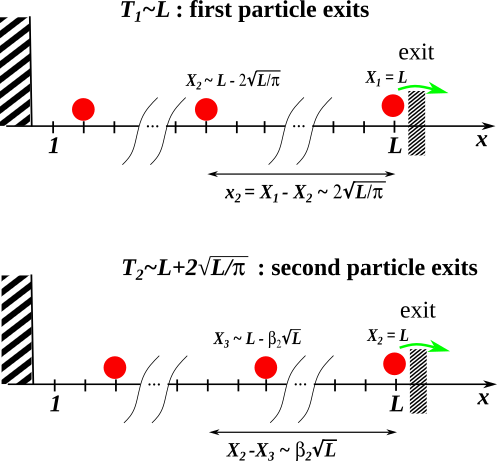}
\caption{Mean exit times of the first two particles. }
\label{fig:exit_times}
\end{figure}
\end{center}
We denote by $x$ the relative position of a particle with respect to the leading one (particle 1), and $X$ its absolute position in the lab frame. Let us first consider the occupancy probability ${\cal P}_2(x;t)$ of the second particle. As we saw, this relative motion is simply described by a continuous diffusion equation. We are then left with the following set of equations in the domain $x<0$ :
\begin{eqnarray} 
& & \partial_t {\cal P}_2(x;t) = \partial_{xx}{\cal P}_2(x;t) \; , \quad x<0 \nonumber \\
& & {\cal P}_2(x;t) \xrightarrow[x \rightarrow -\infty]{} 0 \nonumber \\
& & {\cal P}_2(x;t=0) = \delta(x-0^-)  \\
& & {\cal J}(0;t) \equiv -\partial_{x}{\cal P}_2(x=0^-;t) = 0 \nonumber
\end{eqnarray}
Some comments are in order : here the space domain extends from $x=0$ corresponding to the position of the leading particle up to $x=-\infty$ when the trailing particle stays at rest in the lab frame. At $t=0$ particle 2 is situated next to the leading particle, which in the continuous limit gives the stated initial condition. Finally the exclusion caused by the leading particle is described by a no flux condition ${\cal J}(x=0) =0$. The solution of this set of equations is elementary and is twice the fundamental solution of the $1D$ diffusion equation. This immediately leads to the average position for the second particle with respect to the first one: $\langle x_2(t) \rangle = - 2 \sqrt{t / \pi}$. When the leading particle exits at a mean time $T_1 = L$ the second one therefore sits at a position $\langle X_2(L)\rangle = L-2 \sqrt{L / \pi}$ in the lab frame. Since
we demonstrated the equivalence of the finite and infinite lattice frames for the exit problem, we can assume particle 2 to be then unconstrained. Hence, it needs an additional time $T_2 - T_1 = 2 \sqrt{L / \pi}$ to exit, (see Figure\ref{fig:exit_times} for a pictorial view). 
This  reproduces the result obtained previously by our exact algebraic computations (\ref{eq:TF2L}), and this validates our continuous approach. 

Encouraged by this first result, we seek a recursive scheme to obtain the mean position of the $(n+1)$-th particle with time, assuming an average position $\langle x_{n}(t)\rangle =-\beta_n \sqrt{t}$ of the previous one, always relative to the first particle.
The set of evolution equations for the $(n+1)$-th particle can then be written as:
\begin{eqnarray} 
\label{eq:recurence}
& & \partial_t {\cal P}_{n+1}(x;t) = \partial_{xx}{\cal P}_{n+1}(x;t) \;, \quad x < \langle x_n(t) \rangle  \nonumber \\
& & {\cal P}_{n+1}(x;t) \xrightarrow[x \rightarrow -\infty]{} 0  \nonumber \\
& & {\cal P}_{n+1}(x;t=0) = \delta(x-0^-) \\
& & \left[ \partial_{x} {\cal P}_{n+1}(x;t) +  \langle \dot{x}_n(t)\rangle {\cal P}_{n+1}(x;t) \right]_{x=\langle x_n(t) \rangle}  = 0  \nonumber
\end{eqnarray}
 The last condition can be established for example by imposing $\frac{d}{dt} \int_{-\infty}^{\langle x_n(t) \rangle} {\cal P}_{n+1}(x;t)  dx  = 0$ which results from the normalization condition. It enforces a no flux condition at the moving boundary $x_n(t)$.
The solution of Eqs (\ref{eq:recurence}) is simply 
\begin{equation}
     {\cal P}_{n+1}(x;t) = \frac{1}{ \sqrt{ \pi t} \erfc(\beta_n/2)} e^{-x^2/4t}
\end{equation}
One can check that both the normalization and the no-flux condition, which are related, are satisfied by the above solution \textit{when the boundary is moving $\propto \sqrt{t}$}.

The average velocity $\frac{d}{dt} \langle x_{n+1}(t) \rangle = -\beta_{n+1} /(2 \sqrt{t})$ can finally be calculated by taking the mean of the diffusion equation as $ \frac{d}{dt} \langle x_{n+1}(t) \rangle = - {\cal P}_{n+1}(0;t)$, and this allows us to write the following recursion relation : 
\begin{equation} \label{eq:betarecursion}
 \beta_{n+1}= \frac{2}{\sqrt{\pi}} \frac{ \exp{(-\beta_n^2 /4)} }{ \erfc(\beta_n/2)}
\end{equation}
initiating at $\beta_1 = 0$. This is the central analytical result which allows to calculate the actual average position of the nth particle as a function of time in the lab frame as
$\langle X_n(t)\rangle =t-\beta_n \sqrt{t}$.

\begin{center}
\begin{figure}[h] 
\includegraphics[scale = 0.13]{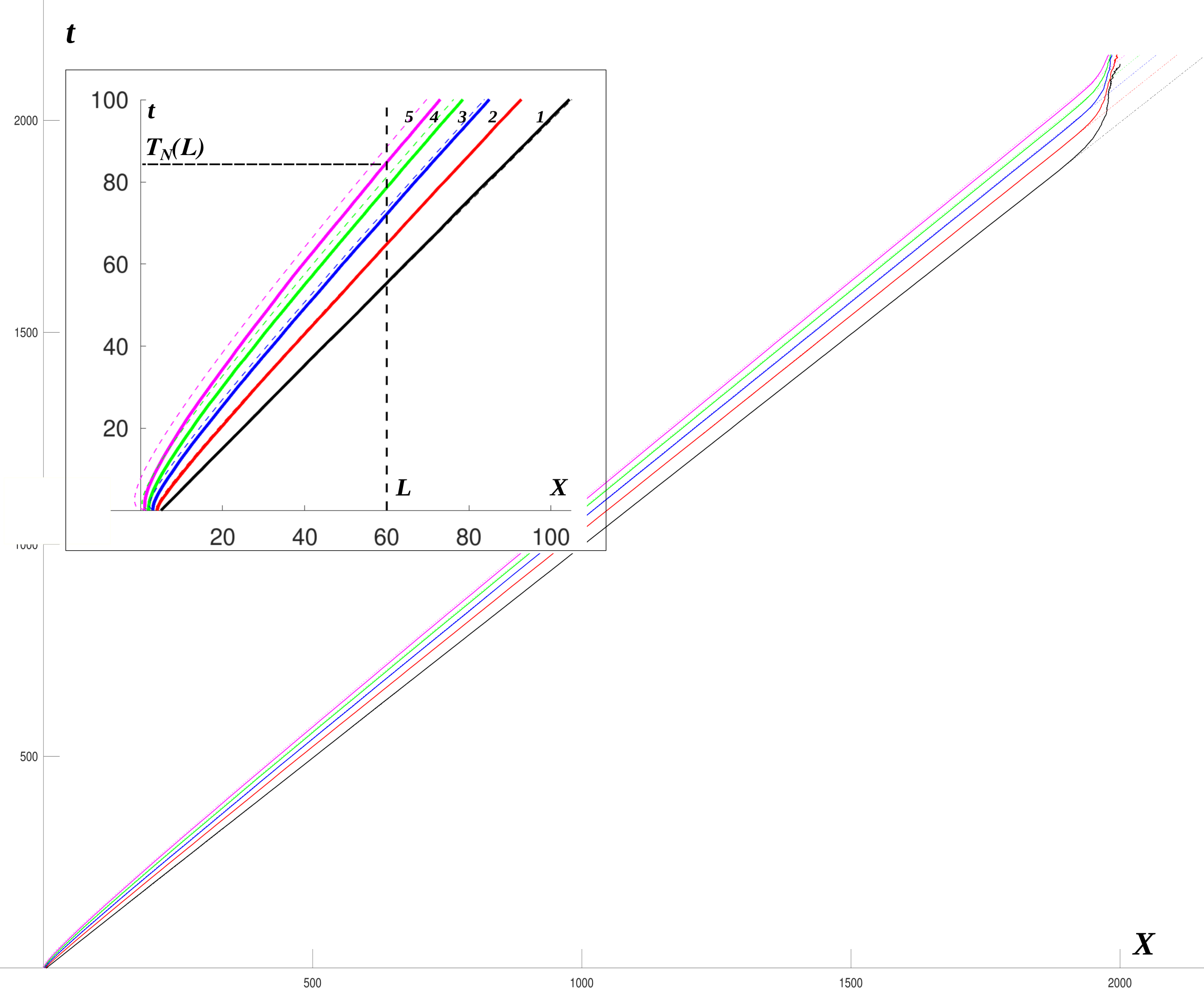}
\caption{Mean trajectories  $\langle X_n(t)\rangle =t-\beta_n \sqrt{t}$ of successive $n=1,2,3,4,5$ TASEP particles from the continuous approach (dotted lines). Inset is a zoom near the origin.
In comparison, the Gillespie-simulated trajectories in solid lines for 1000 different histories show that that the diffusion approach is quite good, see section \ref{Gillespie} (note that the artefact at the end of the simulation near $X=2000$ is due to taking a mean position over a non constant ensemble of particles, since exiting particles are disappearing the simulation at $L=2000$. }
\label{fig:mean_trajectories}
\end{figure}
\end{center}

In Figure \ref{fig:mean_trajectories}, we have plotted the resulting mean trajectories of the first $5$ particles in the lab frame as a function of time, (see dashed lines). For comparison we also plotted in solid lines the mean trajectories obtained by a direct Gillespie-type simulation of the TASEP over 1000 replicas and $L =2000$ sites, see section \ref{Gillespie}. As is clear in the insert, the continuous diffusion approximation breaks down at short times, since TASEP particles can not have negative velocities in the lab frame. It however works remarkably well at larger times/positions, giving the exact large $L$ asymptotic for $N=1$ and 2 and an error of the order of a fraction of percent for small values of $N$ (see below).

Finally, for $N$ particles and large $L$ the asymptotic behavior for the mean exit time is :
\begin{equation} \label{eq:exittime}
 T_N(L) \simeq L + \beta_N \sqrt{L}
\end{equation}
where $\beta_n$ is given by Eq. (\ref{eq:betarecursion}).

\section{Gillespie simulations of exit times} \label{Gillespie}

\begin{center}
\begin{figure}[h]
\includegraphics[scale = 0.27]{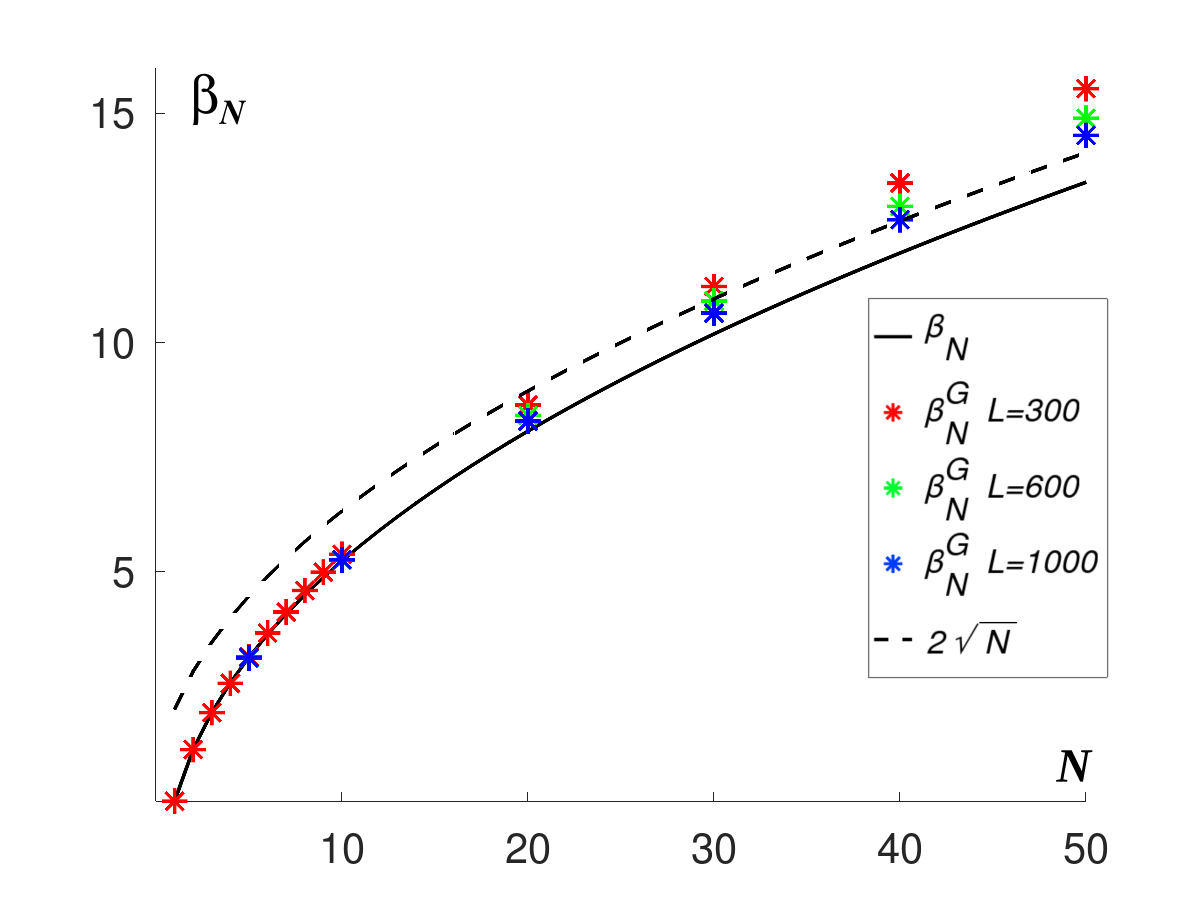}
\caption{Values of $\beta_N^G$ obtained directly by Gillespie simulations (stars : red, green, blue  $L=300, 600, 1000$ respectively) compared to the values of $\beta_N$ obtained by the recursive scheme based on the diffusion equation approach, equation (\ref{eq:betarecursion}) (solid curve). The dashed line correspond to the limit of vanishing densities of reference \cite{Johansson2000}, see section \ref{conclusion} below. }
\label{fig:betaN}
\end{figure}
\end{center}

Numerical simulations were also done to compute directly the exit times. In order to compare our results with real data, we simulated the emptying of a TASEP from an initial step condition using a continuous time Gillespie algorithm, see appendix \ref{Gillespie_details} for details. The simulations were done with $L=300, 600$ and $1000$ sites, and with up to $N=50$ particles. The rather modest number of copies was generally enough to ensure a reasonable error on the mean exit time, since this quantity is in itself a mean of different Gillespie times along one single history. The values of the coefficients $\beta_N^G(L)$
were then computed using the definition 
\begin{equation}
\beta_N^G(L) \equiv \frac{T_N^G(L) - L}{\sqrt{L}} \label{eq:defbeta} 
\end{equation}
and compared to the values obtained by our recursion equation (\ref{eq:betarecursion}) in figure \ref{fig:betaN}, see the red-green-blue stars and black solid curve respectively. 
The agreement between these two independent methods (diffusion approximation calculation and Gillespie simulations) is excellent for $L=1000$ and for small values of $N$, with a vanishing relative error at $N=1$ and 2 since the diffusion method gives the exact result there, and a relative error ranging from +1\% for $N=3$ to -0.7\% for $N=10$. This ultimately validates our diffusion continuous approach. We also note that for large $N$ values, the Gillespie simulations are getting closer to our estimation  (\ref{eq:exittime}) when increasing $L$, with an error of only -7\% for $N=50$ and $L=1000$. Actually we conjecture that 
our diffusion scheme produces an exact result in the limit $L \rightarrow \infty $ and $N\rightarrow \infty $ with $N/L \rightarrow 0$ as we discuss in the next section.

\section{Discussion and conclusions} \label{conclusion}

In this work, we have considered the question of the mean time taken by $N$ particles to empty a lattice with $L$ sites while being transported according to the rules of the TASEP model and starting from the leftmost sites of that lattice (step initial condition). We have investigated two slightly different versions of that problem: A) particles definitively exit the lattice as they leave the site $L$ and B) particles keep moving along an infinite lattice after they have crossed the $L$th site. 
For $N=2$ particles, we have found the exact mean exit time for both problems and we have shown that they have a common asymptotic behaviour at large $L$ equal to  
$T_2(L) =  L + \frac{2}{\sqrt{\pi}}\, \sqrt{L} +{\cal O}(L^{-1/2})$.

Still for $N=2$ particles and within the framework of problem B, we have revisited that result by showing that the probability distribution of the distance between the particles obey a master equation that is a discrete version of a diffusion equation. From there, we have calculated the mean distance as a function of time and rederived the asymptotic behaviour of the mean exit time. Then, generalizing this approach to $N\geq 3$ particles, we have devised an approximate diffusion model for the relative positions of consecutive particles that leads to a mean exit time for $N$ particles that behaves for large $L$ as
$T_N(L) \sim L+\beta_N \sqrt{L}$ where $\beta_N$ can be calculated recursively. Finally, we have confirmed the validity of this approximation by Gillespie simulations for values of $N \ll L$.

Our diffusion model seems to work particularly well for a small finite number of particles. In the limit where the lattice size becomes infinite, $L\to \infty$, the average particle density $N/L$ tends therefore to zero. It is nonetheless tempting to try to extrapolate our results to a number of particles proportional to the lattice size, $N =\mu L$ (as $L \to \infty$) with a proportionality coefficient $\mu \ll 1$ in order to keep $N \ll L$. Assuming equation (\ref{eq:betarecursion}) to be still valid for large values of $N \ll L$,
 the asymptotic behaviour $\beta_N \sim 2 N^{1/2}$ for large $N$ obtained from Eq. \eqref{eq:betarecursion} would yield 
\begin{equation} \label{eq:largeN}
 T_N(L) \simeq L + 2 \sqrt{N L}.
\end{equation}
Letting $N=\mu L$ then provides the following asymptotic behaviour 
\begin{equation}
T_{\mu L}(L) \sim (1+2\sqrt{\mu}) L,\quad (L\to \infty, \mu \ll 1)\, .
\end{equation}
This is to be compared to the exact known result $T_{\mu L}(L) = (1+\sqrt{\mu})^2 L$ obtained by Johansson for $N=\mu L$ in the limit $L \to \infty$ and $\mu$ finite (see ref. \cite{Johansson2000}, theorem 1.6, Eq. (1.19) in which $\gamma = 1/\mu$ \footnote{In this theorem $\gamma$ stands for the proportionality coefficient between the number $M$ of steps undergone by a particle and its rank $N$, $M=\gamma N$. As the $N$th particle is our leftmost particle, it has to hop $L$ times for the lattice to be empty. Therefore $M \equiv L$ and $\gamma = 1/\mu$.}). The corresponding value of $\beta_N^J$ defined as in equation (\ref{eq:defbeta}) reads $\beta_N^J = (2 + \sqrt \mu) \sqrt N$. Using this result in the limit $\mu=0$ corresponding to our vanishing density regime, we have also plotted in Figure \ref{fig:betaN} the corresponding $\beta_N^J = 2 \sqrt N$ (black dashed line). We can see that our approximation gives a much better estimate of $T_L(N)$ in the finite $N$ regime and behaves decently at $N$ large, with the same asymptotic value of $\beta_N$. This leads us to conjecture that equation (\ref{eq:largeN}) is exact in the limit $L \rightarrow \infty $ and $N\rightarrow \infty $ with the density $\mu=N/L \rightarrow 0$, a region outside of the scope of ref. \cite{Johansson2000}.

Another problem of interest is the exit time of $N$ particles transported without exclusion. In that case, particles are all independent. They wait for a time $t$ distributed according to the exponential distribution $e^{-t}$ between two consecutive jumps to the right, may overtake each other and occupy the same lattice site as others. The particle to last exit the lattice among the $N$, irrespective of its initial location, sets the exit time. Evaluating the distribution obeyed by the latter thus simply amounts to finding the distribution of the maximum of the individual exit times of each of the $N$ particles (that depend on their initial location). For $N=2$ particles starting respectively from sites 1 and 2 of an $L$-site lattice, it can be shown that the exit time asymptotically reads
$T_2(L) \sim  L + \frac{1}{\sqrt{\pi}}\sqrt{L} + O(1)$, for large $L$ \footnote{The two first terms of this  asymptotic behaviour would be the same for 2 particles both initially located on site 1 of the lattice.}. Strikingly, we see that the $\sqrt{L}$ correction to the mean exit time of a single particle is not solely attributable to the exclusion effect of the TASEP model. It also occurs in independent particles as a by-product of the distribution of the maximum of their individual exit times, although with a different prefactor (half of the TASEP one for 2 particles). Preliminary analytical and numerical results seem to show that for a large number $N$ of particles, all starting from site 1 of the lattice, the prefactor of the $\sqrt{L}$ correction of the exit time is proportional to $\ln N $ as $N \to \infty$. This behaviour is to be contrasted with the $\sqrt{N}$ correction obtained in presence of exclusion for the TASEP model. 

Finally, this study can be seen as a step towards the calculation of exit times of some more refined transport models. For example, one could try to test the diffusion approximation used in this paper to compute probabilities of interest studied in the clearance problem of \cite{Cividini_2017}.
Queuing problems \cite{Arita2009} or experimental microfluidic setups \cite{PhysRevE.91.022109} could also benefit from our approach (e.g by relaxing the exclusion constraint for the queuing problem, or allowing for bidirectional transport like in the ASEP or SEP models, see\cite{Derrida2009SSEP, NAGAO2004487}).

\appendix

\section{Some exact results for small lattices} \label{ERSL}
In table \ref{tab:ex}, we list some exact results for the exit time distributions and their MET that can be obtained from the method exposed in section \ref{ETDsection}. As may readily be checked from the third column of this table, the Mean Exit time of $N=2$ particles on a finite lattice $\{1,L\}$ ($L \leq 10$) agrees with the exact formula provided in \eqref{MET2partf}. Laplace transforms of the time distributions have been given up to $L=5$ only for they then become somewhat lengthy. From $L\geq 3$ onwards, the denominator of $\tilde{p}_{2,L}(s)$ is $(s+1)^{L+1}(s+2)^{2L-5}$. The constant coefficient of the numerator polynomial is $2^{2L-5}$ and its highest degree coefficient ($s^{L-3}$ for $L \geq 3$) is the Catalan number $C(L)=\binom{2L}{L}/(L+1)$ (valid for $L\geq 2$).
As for $\tilde{p}_{3,L}(s)$, its denominator is given by 
$(s+1)^{L+2}(s+2)^{2L-3}(s+3)^{3L-14}$ for $L\geq 5$.

Exact results for small lattices (up to $L=20$) with $N=2,3$ particles are typically obtained by Maple on a basic laptop within less than a minute computation time.  
These results may serve as benchmarks for simulations. 
\begin{table*}[t]
\caption{\label{tab:ex} Exact Laplace transform and Mean Exit Time of the exit time distribution of a finite lattice $ \{1,L\}$ for $L \leq 10$ and $N=2,3$.}
\begin{tabular*}{\textwidth}{c@{\extracolsep{\fill}}cccc} 
 $L$ & $\tilde{p}_{2,L}(s)$ & $T^{(f)}_2(L)$ & $\tilde{p}_{3,L}(s)$ & $T^{(f)}_3(L)$ \\ 
 \hline
 2 & $\displaystyle \frac{1}{(s+1)^3}$ & $3$ & n.a & n.a \\ 
 3 & $\displaystyle \frac{2}{(s+1)^4(s+2)}$ & $\displaystyle \frac{9}{2}$ & $\displaystyle \frac{2}{(s+1)^5(s+2)}$ & $\displaystyle \frac{11}{2}$  \\
 4 & $\displaystyle \frac{5s+8}{(s+1)^5(s+2)^3}$ & $\displaystyle \frac{47}{8}$ & $\displaystyle \frac{12s^2 + 39s+32}{(s+1)^6(s+2)^5}$ & $\displaystyle \frac{233}{32}$  \\
 5 & $\displaystyle \frac{2(7s^2+21s+16)}{(s+1)^6(s+2)^5}$ & $\displaystyle \frac{115}{16}$ &  $\displaystyle \frac{110s^3 + 495s^2 + 751s + 384}{(s+1)^7(s+2)^7(s+3)}$ & $\displaystyle \frac{3409}{384}$  \\
 6 &  $\vdots$ & $\displaystyle \frac{1083}{128}$ &  $\vdots$ & $\displaystyle \frac{107617}{10368} $   \\ 
 7 &  $\vdots$ & $\displaystyle \frac{2485}{256}$ & $\vdots$ & $\displaystyle \frac{13237775}{1119744}$ \\  
 8 &  $\vdots$  & $\displaystyle \frac{11195}{1024}$  & $\vdots$ & $\displaystyle \frac{2132010983}{161243136}$ \\
 9 &  $\vdots$  & $\displaystyle \frac{24867}{2048}$ & $\vdots$ & $\displaystyle \frac{254084494957}{17414258688}$ \\ 
 10 &  $\vdots$  & $\displaystyle \frac{437075}{32768}$ & $\vdots$ & $\displaystyle \frac{7491745364599}{470184984576}$ 
\end{tabular*}
\end{table*}

\section{Exact MET for 2 particles, finite lattice} \label{MET2partfinite}
To find the Mean Exit time of two particles initially located on site 1 and 2 of the finite lattice $\{1,L\}$, it is sufficient, according to Eqs. \eqref{pext} and \eqref{MET}, to find the Laplace Transform (LT) of the probability that particle 2 is on site $L$ of that lattice while particle 1 has left it. We shall denote that quantity by $\tilde{P}_o(L;s)$ where the subscript $o$ indicates that particle 1 has left the lattice. We shall denote by $\tilde{P}(k_2,k_1;s)$ the LT of the probability that particle 1 is at site $k_1$ and particle 2 at site $k_2$ with $1 \leq k_2 < k_1 \leq L$.  Let us write the master equation for the LT $\tilde{P}_o(n;s)$,
$n \in [\![1,L]\!]$. Dropping the $s$ dependence for simplicity, one obtains,
\begin{eqnarray}
   s \tilde{P}_o(1) &=& - \tilde{P}_o(1) + \tilde{P}(1,L) \nonumber \\ 
   s \tilde{P}_o(k) &=& - \tilde{P}_o(k) + \tilde{P}_o(k-1) + \tilde{P}(k,L) \nonumber \\ 
   s\tilde{P}_o(L) &=& - \tilde{P}_o(L) + \tilde{P}_o(L-1)
\end{eqnarray}
where $k \in [\![2,L-1]\!]$. Solving for $\tilde{P}_o(L;s)$
yields
\begin{equation} \label{LTPoLs}
  \tilde{P}_o(L;s)= \sum_{n=1}^{L-1} \frac{\tilde{P}(n,L;s)}{(s+1)^{L-n+1}} 
\end{equation}
We shall now take advantage of the fact that $P(n,L;t)$ is known exactly for it is the probability that two particles located on sites 1 and 2 at $t=0$ be located at site $n$ and $L$, respectively, at time $t$. This transition probability is provided by Schütz in \cite{Schutz1997} who has solved this problem on an infinite lattice. Yet, as none of the particles have left the section $[1,L]$, this probability is exactly the same as for the finite lattice $\{1,L\}$. This makes the necessary connection between the finite and infinite lattice problems.    
According to \cite{Schutz1997}, we have 
\begin{equation}
 P(n,L;t) = \begin{vmatrix}
F_0(n-1;t) & F_{-1}(n-2;t) \\
F_1(L-1;t) & F_0(L-2;t) 
\end{vmatrix} \, ,   
\end{equation}
where 
$$F_0(k;t)=\frac{t^k}{k!}e^{-t}\ \textrm{and}\ F_1(k;t)=1 - e^{-t}\sum_{q=0}^{k-1}\frac{t^q}{q!}$$
and where $F_{-1}(k;t)=F_{0}(k;t)-F_{0}(k+1;t)$.
Expanding $P(n,L;t)$ in sums of products of exponentials and powers in $t$ makes it easy to obtain its Laplace transform $\tilde{P}(n,L;s)$. Reinstating the latter in Eq. \eqref{LTPoLs} and using 
\begin{equation}
  T^{(f)}_2(L) = -\left. \frac{d\tilde{P}_o(L;s)}{ds}\right|_{s=0} \, , 
\end{equation}
eventually yields, after a somewhat lengthy calculation, 
\begin{equation}
T^{(f)}_2(L)= L + \frac{L-1}{4^{L-2} } \times \binom{2L-3}{L-1} .
\end{equation}

\section{Exact MET for 2 particles, infinite lattice} \label{MET2partinfinite}
The easiest way to obtain the exact Mean Exit Time (MET) for 2 particles leaving a section $[1,L]$ of an infinite lattice while being initially located on sites 1 and 2 of that section is probably to use the integral formula given by Rakos and Schütz \cite{Rakos2005} for the probability that the second leftmost particle of two initially side by side particles has carried out at least $L$ steps to the right at time $t$. This probability, that is exactly the probability that the two particles have left the section $[1,L]$ by time $t$, is given by 
\begin{equation} \label{PMNtRakos}
    P(L,2,t) = Z \int_{[0,t]^2} \!\!\!\!\!\!\!\! dx_1 dx_2 (x_1x_2)^{L-2} e^{-(x_1+x_2)}(x_1-x_2)^2  
\end{equation}
where
\begin{equation}
    Z = \frac{L-1}{2\left[(L-1)!\right]^2}\, .
\end{equation}
From Eq. \eqref{PMNtRakos}, the corresponding exit time distribution is given by $p_{2,L}(t)=\dot{P}(L,2,t)$, whence the MET
\begin{equation} \label{METT2LRakos1}
    T_2(L) = \int_0^{\infty} \! t \dot{P}(L,2,t) \, dt\, .
\end{equation}
The probability $P(L,2,t)$ may be evaluated in terms of incomplete
gamma functions $\gamma(n,t)=\int_0^t x^{n-1}e^{-x} dx$ as
\begin{equation} \label{PMNtRakos2}
    P(L,2,t) = Z \left[\gamma(L+1,t)\gamma(L-1,t)-\gamma(L,t)^2\right] \, .
\end{equation}
Using this expression and \eqref{METT2LRakos1}, $T_2(L)$ can eventually be cast into the simple form
\begin{equation} \label{METT2LRakos2}
    T_2(L) = L + \frac{2}{\sqrt{\pi}}\frac{\Gamma(L+\frac{1}{2})}{\Gamma(L)}\, ,
\end{equation}
where $\Gamma(L)=\gamma(L,\infty)$ is the complete gamma function.
If we compare the asymptotic expressions of $T_2(L)$ and $T^{(f)}_2(L)$ (the MET of two particles leaving a finite lattice $\{1,L\}$ - see Eq. \eqref{MET2partf}), we find that they differ at order $L^{-1/2}$. More precisely,
\begin{equation} \label{MET2partdiff}
    T_2(L) - T^{(f)}_2(L) = \frac{1}{\sqrt{\pi L}} + {\cal O}\left(L^{-3/2}\right)\, .
\end{equation}
As expected, the time needed by the two particles to exit the section $[1,L]$ of an infinite lattice is slightly longer than the time needed to exit the finite lattice $\{1,L\}$ given that when the rightmost particle has gone out of $\{1,L\}$, the last one is free to move ahead while it can still be hindered by the front particle on the infinite lattice.

\section{Exact mean relative distance for 2 particles} \label{MRD2partinfinite}
To obtain the mean relative distance between two particles, initially side by side, on an infinite lattice, we first take the Laplace transform of Eqs. \eqref{eq:Masterrelativepart} and \eqref{eq:discrete_noflux}: 
\begin{eqnarray}
 && s\tilde{\cal P}(1;s)-1 = -\tilde{\cal P}(1;s)+\tilde{\cal P}(2;s) \nonumber \\ 
 && s\tilde{\cal P}(\chi;s) = -2 \tilde{\cal P}(\chi;s) + \tilde{\cal P}(\chi+1;s) + \tilde{\cal P}(\chi-1;s),\nonumber 
\end{eqnarray}
where $\chi\geq 2 $. Solving for $\tilde{\cal P}(\chi;s)$ and taking into account the fact that $\sum_{\chi\geq 1}{\cal P}(\chi;t)=1$ (i.e. $\sum_{\chi\geq 1}\tilde{\cal P}(\chi;s)=1/s$), we obtain
\begin{equation}
 \tilde{\cal P}(\chi;s)=   \frac{1-\lambda}{s}\, \lambda^{\chi-1} ,
\end{equation}
where
\begin{equation}
\lambda = 1+\frac{s}{2} -\sqrt{\left(1+\frac{s}{2}\right)^2-1} \,.
\end{equation}
Then,
\begin{equation}
    \langle \tilde{\chi}(s) \rangle := \sum_{\chi\geq 1}\chi \tilde{\cal P}(\chi;s) = \frac{1}{s(1-\lambda)}\, ,
\end{equation}
and, upon inverting that expression, we finally obtain
\begin{equation} 
    \langle \chi(t) \rangle = \frac{e^{-2t}}{2}[(4t+1)I_0(2t)+4tI_1(2t)] +\frac{1}{2}
\end{equation}
where $I_k$ is the modified Bessel function of order $k$: $I_0(x)=\sum_{n \geq 0} x^{2n}/[4^n(n!)^2]$ and $I_1(x)=x I_0'(x)$ where the prime denotes derivatives wrt $x$.
\vspace{10pt}

\section{Gillespie simulations details} \label{Gillespie_details}
Numerical simulation were performed using Octave on a DELL XPS13. The continuous time Gillespie method was used in order to produce an in-silico realization of equation \ref{METASEP}. In this method, each history simulate a stochastic trajectory associated with the TASEP master equation. Most of the simulations were done using $10^3$ histories, in order to keep the computation time manageable on a laptop, especially for large values of $N \simeq 50$ and $L\simeq 1000$. To estimate the error for mean values such as  $T^G_N(L)$, we performed 20 independent simulations of $10^3$ histories and obtained a dispersion of values of the order of $\Delta T \sim 0.5$ for $T^G_N(L) \sim 300-1000$. The same procedure was then used with $10^4$ histories and, as expected, lowered this figure to $\Delta T \sim 0.15$.

The precision obtained with $10^3$ copies was usually enough to compare the simulations results with our theoretical value $\beta_N$. For small $N < 5$ however, the values of $\beta_N(L)$ estimated by the different methods and for different length $L=300-1000$ are very close, and it was necessary to use $10^4$ copies to order the different values of $\beta_N(L)$ properly. It was found that :
\begin{itemize}
    \item for fixed $N$ the values of $\beta_N^G(L)$ are systematically decreasing when increasing $L$, as seen in Figure \ref{fig:betaN}.
    \item for values of $N>10$ our result (\ref{eq:betarecursion}) underestimates the value of the coefficient, $\beta_N(L)$, while for small values it is overestimating.
    \item for $N \le 10$ the relative error of equation (\ref{eq:betarecursion}) with respect to our best estimate of the exact $\beta_N$, obtained with the highest copy number and the highest $L$, is less than 1\%, (0 for $N=1$ and 2 since our expression is then exact), and reaches -7 \% for $N=50$.
\end{itemize}

\bibliography{biblio.bib}

\begin{thebibliography}{33}%
\makeatletter
\providecommand \@ifxundefined [1]{%
 \@ifx{#1\undefined}
}%
\providecommand \@ifnum [1]{%
 \ifnum #1\expandafter \@firstoftwo
 \else \expandafter \@secondoftwo
 \fi
}%
\providecommand \@ifx [1]{%
 \ifx #1\expandafter \@firstoftwo
 \else \expandafter \@secondoftwo
 \fi
}%
\providecommand \natexlab [1]{#1}%
\providecommand \enquote  [1]{``#1''}%
\providecommand \bibnamefont  [1]{#1}%
\providecommand \bibfnamefont [1]{#1}%
\providecommand \citenamefont [1]{#1}%
\providecommand \href@noop [0]{\@secondoftwo}%
\providecommand \href [0]{\begingroup \@sanitize@url \@href}%
\providecommand \@href[1]{\@@startlink{#1}\@@href}%
\providecommand \@@href[1]{\endgroup#1\@@endlink}%
\providecommand \@sanitize@url [0]{\catcode `\\12\catcode `\$12\catcode
  `\&12\catcode `\#12\catcode `\^12\catcode `\_12\catcode `\%12\relax}%
\providecommand \@@startlink[1]{}%
\providecommand \@@endlink[0]{}%
\providecommand \url  [0]{\begingroup\@sanitize@url \@url }%
\providecommand \@url [1]{\endgroup\@href {#1}{\urlprefix }}%
\providecommand \urlprefix  [0]{URL }%
\providecommand \Eprint [0]{\href }%
\providecommand \doibase [0]{https://doi.org/}%
\providecommand \selectlanguage [0]{\@gobble}%
\providecommand \bibinfo  [0]{\@secondoftwo}%
\providecommand \bibfield  [0]{\@secondoftwo}%
\providecommand \translation [1]{[#1]}%
\providecommand \BibitemOpen [0]{}%
\providecommand \bibitemStop [0]{}%
\providecommand \bibitemNoStop [0]{.\EOS\space}%
\providecommand \EOS [0]{\spacefactor3000\relax}%
\providecommand \BibitemShut  [1]{\csname bibitem#1\endcsname}%
\let\auto@bib@innerbib\@empty
\bibitem [{\citenamefont {Johansson}(2000)}]{Johansson2000}%
  \BibitemOpen
  \bibfield  {author} {\bibinfo {author} {\bibfnamefont {K.}~\bibnamefont
  {Johansson}},\ }\href {https://doi.org/10.1007/s002200050027} {\bibfield
  {journal} {\bibinfo  {journal} {Communications in Mathematical Physics}\
  }\textbf {\bibinfo {volume} {209}},\ \bibinfo {pages} {437} (\bibinfo {year}
  {2000})}\BibitemShut {NoStop}%
\bibitem [{\citenamefont {MacDonald}\ \emph {et~al.}(1968)\citenamefont
  {MacDonald}, \citenamefont {Gibbs},\ and\ \citenamefont
  {Pipkin}}]{MacDonald1968}%
  \BibitemOpen
  \bibfield  {author} {\bibinfo {author} {\bibfnamefont {C.~T.}\ \bibnamefont
  {MacDonald}}, \bibinfo {author} {\bibfnamefont {J.~H.}\ \bibnamefont
  {Gibbs}},\ and\ \bibinfo {author} {\bibfnamefont {A.~C.}\ \bibnamefont
  {Pipkin}},\ }\href {https://doi.org/10.1002/bip.1968.360060102} {\bibfield
  {journal} {\bibinfo  {journal} {Biopolymers}\ }\textbf {\bibinfo {volume}
  {6}},\ \bibinfo {pages} {1} (\bibinfo {year} {1968})}\BibitemShut {NoStop}%
\bibitem [{\citenamefont {Schadschneider}\ \emph {et~al.}(2011)\citenamefont
  {Schadschneider}, \citenamefont {Chowdhury},\ and\ \citenamefont
  {Nishinari}}]{Schadschneider:1564539}%
  \BibitemOpen
  \bibfield  {author} {\bibinfo {author} {\bibfnamefont {A.}~\bibnamefont
  {Schadschneider}}, \bibinfo {author} {\bibfnamefont {D.}~\bibnamefont
  {Chowdhury}},\ and\ \bibinfo {author} {\bibfnamefont {K.}~\bibnamefont
  {Nishinari}},\ }\href {https://cds.cern.ch/record/1564539} {\emph {\bibinfo
  {title} {Stochastic transport in complex systems: from molecules to
  vehicles}}}\ (\bibinfo  {publisher} {Elsevier},\ \bibinfo {address}
  {Amsterdam},\ \bibinfo {year} {2011})\BibitemShut {NoStop}%
\bibitem [{\citenamefont {Arita}(2009)}]{Arita2009}%
  \BibitemOpen
  \bibfield  {author} {\bibinfo {author} {\bibfnamefont {C.}~\bibnamefont
  {Arita}},\ }\href {https://doi.org/10.1103/PhysRevE.80.051119} {\bibfield
  {journal} {\bibinfo  {journal} {Phys. Rev. E}\ }\textbf {\bibinfo {volume}
  {80}},\ \bibinfo {pages} {051119} (\bibinfo {year} {2009})}\BibitemShut
  {NoStop}%
\bibitem [{\citenamefont {Arita}\ and\ \citenamefont
  {Schadschneider}(2014)}]{ARITA2014}%
  \BibitemOpen
  \bibfield  {author} {\bibinfo {author} {\bibfnamefont {C.}~\bibnamefont
  {Arita}}\ and\ \bibinfo {author} {\bibfnamefont {A.}~\bibnamefont
  {Schadschneider}},\ }\href
  {https://doi.org/https://doi.org/10.1016/j.trpro.2014.09.012} {\bibfield
  {journal} {\bibinfo  {journal} {Transportation Research Procedia}\ }\textbf
  {\bibinfo {volume} {2}},\ \bibinfo {pages} {87} (\bibinfo {year} {2014})},\
  \bibinfo {note} {the Conference on Pedestrian and Evacuation Dynamics 2014
  (PED 2014), 22-24 October 2014, Delft, The Netherlands}\BibitemShut {NoStop}%
\bibitem [{\citenamefont {Sasamoto}(2007)}]{Sasamoto_2007}%
  \BibitemOpen
  \bibfield  {author} {\bibinfo {author} {\bibfnamefont {T.}~\bibnamefont
  {Sasamoto}},\ }\href {https://doi.org/10.1088/1742-5468/2007/07/P07007}
  {\bibfield  {journal} {\bibinfo  {journal} {Journal of Statistical Mechanics:
  Theory and Experiment}\ }\textbf {\bibinfo {volume} {2007}},\ \bibinfo
  {pages} {P07007} (\bibinfo {year} {2007})}\BibitemShut {NoStop}%
\bibitem [{\citenamefont {Chowdhury}\ \emph {et~al.}(2000)\citenamefont
  {Chowdhury}, \citenamefont {Santen},\ and\ \citenamefont
  {Schadschneider}}]{CHOWDHURY2000199}%
  \BibitemOpen
  \bibfield  {author} {\bibinfo {author} {\bibfnamefont {D.}~\bibnamefont
  {Chowdhury}}, \bibinfo {author} {\bibfnamefont {L.}~\bibnamefont {Santen}},\
  and\ \bibinfo {author} {\bibfnamefont {A.}~\bibnamefont {Schadschneider}},\
  }\href {https://doi.org/https://doi.org/10.1016/S0370-1573(99)00117-9}
  {\bibfield  {journal} {\bibinfo  {journal} {Physics Reports}\ }\textbf
  {\bibinfo {volume} {329}},\ \bibinfo {pages} {199} (\bibinfo {year}
  {2000})}\BibitemShut {NoStop}%
\bibitem [{\citenamefont {Zia}\ \emph {et~al.}(2011)\citenamefont {Zia},
  \citenamefont {Dong},\ and\ \citenamefont {Schmittmann}}]{Zia2011}%
  \BibitemOpen
  \bibfield  {author} {\bibinfo {author} {\bibfnamefont {R.}~\bibnamefont
  {Zia}}, \bibinfo {author} {\bibfnamefont {J.}~\bibnamefont {Dong}},\ and\
  \bibinfo {author} {\bibfnamefont {B.}~\bibnamefont {Schmittmann}},\ }\href
  {https://doi.org/10.1007/s10955-011-0183-1} {\bibfield  {journal} {\bibinfo
  {journal} {Journal of Statistical Physics}\ }\textbf {\bibinfo {volume}
  {144}},\ \bibinfo {pages} {405} (\bibinfo {year} {2011})}\BibitemShut
  {NoStop}%
\bibitem [{\citenamefont {Derrida}\ \emph {et~al.}(1992)\citenamefont
  {Derrida}, \citenamefont {Domany},\ and\ \citenamefont
  {Mukamel}}]{Derrida1992}%
  \BibitemOpen
  \bibfield  {author} {\bibinfo {author} {\bibfnamefont {B.}~\bibnamefont
  {Derrida}}, \bibinfo {author} {\bibfnamefont {E.}~\bibnamefont {Domany}},\
  and\ \bibinfo {author} {\bibfnamefont {D.}~\bibnamefont {Mukamel}},\ }\href
  {https://doi.org/10.1007/bf01050430} {\bibfield  {journal} {\bibinfo
  {journal} {Journal of Statistical Physics}\ }\textbf {\bibinfo {volume}
  {69}},\ \bibinfo {pages} {667} (\bibinfo {year} {1992})}\BibitemShut
  {NoStop}%
\bibitem [{\citenamefont {Derrida}\ \emph {et~al.}(1993)\citenamefont
  {Derrida}, \citenamefont {Evans}, \citenamefont {Hakim},\ and\ \citenamefont
  {Pasquier}}]{Derrida1993DE}%
  \BibitemOpen
  \bibfield  {author} {\bibinfo {author} {\bibfnamefont {B.}~\bibnamefont
  {Derrida}}, \bibinfo {author} {\bibfnamefont {M.~R.}\ \bibnamefont {Evans}},
  \bibinfo {author} {\bibfnamefont {V.}~\bibnamefont {Hakim}},\ and\ \bibinfo
  {author} {\bibfnamefont {V.}~\bibnamefont {Pasquier}},\ }\href
  {https://doi.org/10.1088/0305-4470/26/7/011} {\bibfield  {journal} {\bibinfo
  {journal} {Journal of Physics A: Mathematical and General}\ }\textbf
  {\bibinfo {volume} {26}},\ \bibinfo {pages} {1493} (\bibinfo {year}
  {1993})}\BibitemShut {NoStop}%
\bibitem [{\citenamefont {Lazarescu}\ and\ \citenamefont
  {Mallick}(2011)}]{Lazarescu2011}%
  \BibitemOpen
  \bibfield  {author} {\bibinfo {author} {\bibfnamefont {A.}~\bibnamefont
  {Lazarescu}}\ and\ \bibinfo {author} {\bibfnamefont {K.}~\bibnamefont
  {Mallick}},\ }\href {https://doi.org/10.1088/1751-8113/44/31/315001}
  {\bibfield  {journal} {\bibinfo  {journal} {Journal of Physics A:
  Mathematical and Theoretical}\ }\textbf {\bibinfo {volume} {44}},\ \bibinfo
  {pages} {315001} (\bibinfo {year} {2011})}\BibitemShut {NoStop}%
\bibitem [{\citenamefont {Derrida}(1998)}]{DERRIDA199865}%
  \BibitemOpen
  \bibfield  {author} {\bibinfo {author} {\bibfnamefont {B.}~\bibnamefont
  {Derrida}},\ }\href
  {https://doi.org/https://doi.org/10.1016/S0370-1573(98)00006-4} {\bibfield
  {journal} {\bibinfo  {journal} {Physics Reports}\ }\textbf {\bibinfo {volume}
  {301}},\ \bibinfo {pages} {65} (\bibinfo {year} {1998})}\BibitemShut
  {NoStop}%
\bibitem [{\citenamefont {Chou}\ \emph {et~al.}(2011)\citenamefont {Chou},
  \citenamefont {Mallick},\ and\ \citenamefont {Zia}}]{Chou2011}%
  \BibitemOpen
  \bibfield  {author} {\bibinfo {author} {\bibfnamefont {T.}~\bibnamefont
  {Chou}}, \bibinfo {author} {\bibfnamefont {K.}~\bibnamefont {Mallick}},\ and\
  \bibinfo {author} {\bibfnamefont {R.~K.~P.}\ \bibnamefont {Zia}},\ }\href
  {https://doi.org/10.1088/0034-4885/74/11/116601} {\bibfield  {journal}
  {\bibinfo  {journal} {Reports on Progress in Physics}\ }\textbf {\bibinfo
  {volume} {74}},\ \bibinfo {pages} {116601} (\bibinfo {year}
  {2011})}\BibitemShut {NoStop}%
\bibitem [{\citenamefont {Schütz}(1997)}]{Schutz1997}%
  \BibitemOpen
  \bibfield  {author} {\bibinfo {author} {\bibfnamefont {G.}~\bibnamefont
  {Schütz}},\ }\href {https://doi.org/10.1007/BF02508478} {\bibfield
  {journal} {\bibinfo  {journal} {Journal of Statistical Physics}\ }\textbf
  {\bibinfo {volume} {88}},\ \bibinfo {pages} {427} (\bibinfo {year}
  {1997})}\BibitemShut {NoStop}%
\bibitem [{\citenamefont {Rákos}\ and\ \citenamefont
  {Schütz}(2005)}]{Rakos2005}%
  \BibitemOpen
  \bibfield  {author} {\bibinfo {author} {\bibfnamefont {A.}~\bibnamefont
  {Rákos}}\ and\ \bibinfo {author} {\bibfnamefont {G.}~\bibnamefont
  {Schütz}},\ }\href {https://doi.org/10.1007/s10955-004-8819-z} {\bibfield
  {journal} {\bibinfo  {journal} {Journal of Statistical Physics}\ }\textbf
  {\bibinfo {volume} {118}},\ \bibinfo {pages} {511} (\bibinfo {year}
  {2005})}\BibitemShut {NoStop}%
\bibitem [{\citenamefont {Nagao}\ and\ \citenamefont
  {Sasamoto}(2004)}]{NAGAO2004487}%
  \BibitemOpen
  \bibfield  {author} {\bibinfo {author} {\bibfnamefont {T.}~\bibnamefont
  {Nagao}}\ and\ \bibinfo {author} {\bibfnamefont {T.}~\bibnamefont
  {Sasamoto}},\ }\href
  {https://doi.org/https://doi.org/10.1016/j.nuclphysb.2004.08.016} {\bibfield
  {journal} {\bibinfo  {journal} {Nuclear Physics B}\ }\textbf {\bibinfo
  {volume} {699}},\ \bibinfo {pages} {487} (\bibinfo {year}
  {2004})}\BibitemShut {NoStop}%
\bibitem [{\citenamefont {Grabsch}\ \emph {et~al.}(2023)\citenamefont
  {Grabsch}, \citenamefont {Rizkallah}, \citenamefont {Poncet}, \citenamefont
  {Illien},\ and\ \citenamefont {B\'enichou}}]{Grabsch2023}%
  \BibitemOpen
  \bibfield  {author} {\bibinfo {author} {\bibfnamefont {A.}~\bibnamefont
  {Grabsch}}, \bibinfo {author} {\bibfnamefont {P.}~\bibnamefont {Rizkallah}},
  \bibinfo {author} {\bibfnamefont {A.}~\bibnamefont {Poncet}}, \bibinfo
  {author} {\bibfnamefont {P.}~\bibnamefont {Illien}},\ and\ \bibinfo {author}
  {\bibfnamefont {O.}~\bibnamefont {B\'enichou}},\ }\href
  {https://doi.org/10.1103/PhysRevE.107.044131} {\bibfield  {journal} {\bibinfo
   {journal} {Phys. Rev. E}\ }\textbf {\bibinfo {volume} {107}},\ \bibinfo
  {pages} {044131} (\bibinfo {year} {2023})}\BibitemShut {NoStop}%
\bibitem [{\citenamefont {Krapivsky}\ \emph {et~al.}(2014)\citenamefont
  {Krapivsky}, \citenamefont {Mallick},\ and\ \citenamefont
  {Sadhu}}]{PhysRevLett.113.078101}%
  \BibitemOpen
  \bibfield  {author} {\bibinfo {author} {\bibfnamefont {P.~L.}\ \bibnamefont
  {Krapivsky}}, \bibinfo {author} {\bibfnamefont {K.}~\bibnamefont {Mallick}},\
  and\ \bibinfo {author} {\bibfnamefont {T.}~\bibnamefont {Sadhu}},\ }\href
  {https://doi.org/10.1103/PhysRevLett.113.078101} {\bibfield  {journal}
  {\bibinfo  {journal} {Phys. Rev. Lett.}\ }\textbf {\bibinfo {volume} {113}},\
  \bibinfo {pages} {078101} (\bibinfo {year} {2014})}\BibitemShut {NoStop}%
\bibitem [{\citenamefont {Sanders}\ and\ \citenamefont
  {Ambjörnsson}(2012)}]{Sanders2012}%
  \BibitemOpen
  \bibfield  {author} {\bibinfo {author} {\bibfnamefont {L.~P.}\ \bibnamefont
  {Sanders}}\ and\ \bibinfo {author} {\bibfnamefont {T.}~\bibnamefont
  {Ambjörnsson}},\ }\href {https://doi.org/10.1063/1.4707349} {\bibfield
  {journal} {\bibinfo  {journal} {The Journal of Chemical Physics}\ }\textbf
  {\bibinfo {volume} {136}},\ \bibinfo {pages} {175103} (\bibinfo {year}
  {2012})}\BibitemShut {NoStop}%
\bibitem [{\citenamefont {Sliusarenko}\ \emph {et~al.}(2010)\citenamefont
  {Sliusarenko}, \citenamefont {Gonchar}, \citenamefont {Chechkin},
  \citenamefont {Sokolov},\ and\ \citenamefont {Metzler}}]{Sliusarenko2010}%
  \BibitemOpen
  \bibfield  {author} {\bibinfo {author} {\bibfnamefont {O.~Y.}\ \bibnamefont
  {Sliusarenko}}, \bibinfo {author} {\bibfnamefont {V.~Y.}\ \bibnamefont
  {Gonchar}}, \bibinfo {author} {\bibfnamefont {A.~V.}\ \bibnamefont
  {Chechkin}}, \bibinfo {author} {\bibfnamefont {I.~M.}\ \bibnamefont
  {Sokolov}},\ and\ \bibinfo {author} {\bibfnamefont {R.}~\bibnamefont
  {Metzler}},\ }\href {https://doi.org/10.1103/PhysRevE.81.041119} {\bibfield
  {journal} {\bibinfo  {journal} {Phys. Rev. E}\ }\textbf {\bibinfo {volume}
  {81}},\ \bibinfo {pages} {041119} (\bibinfo {year} {2010})}\BibitemShut
  {NoStop}%
\bibitem [{\citenamefont {Molchan}(1999)}]{Molchan1999}%
  \BibitemOpen
  \bibfield  {author} {\bibinfo {author} {\bibfnamefont {G.~M.}\ \bibnamefont
  {Molchan}},\ }\href {https://doi.org/10.1007/s002200050669} {\bibfield
  {journal} {\bibinfo  {journal} {Communications in Mathematical Physics}\
  }\textbf {\bibinfo {volume} {205}},\ \bibinfo {pages} {97} (\bibinfo {year}
  {1999})}\BibitemShut {NoStop}%
\bibitem [{\citenamefont {Derrida}\ and\ \citenamefont
  {Gerschenfeld}(2009)}]{Derrida2009SSEP}%
  \BibitemOpen
  \bibfield  {author} {\bibinfo {author} {\bibfnamefont {B.}~\bibnamefont
  {Derrida}}\ and\ \bibinfo {author} {\bibfnamefont {A.}~\bibnamefont
  {Gerschenfeld}},\ }\href {https://doi.org/10.1007/s10955-009-9830-1}
  {\bibfield  {journal} {\bibinfo  {journal} {Journal of Statistical Physics}\
  }\textbf {\bibinfo {volume} {137}},\ \bibinfo {pages} {978} (\bibinfo {year}
  {2009})}\BibitemShut {NoStop}%
\bibitem [{\citenamefont {Gonçalves}\ and\ \citenamefont
  {Jara}(2008)}]{Gonçalves2008}%
  \BibitemOpen
  \bibfield  {author} {\bibinfo {author} {\bibfnamefont {P.}~\bibnamefont
  {Gonçalves}}\ and\ \bibinfo {author} {\bibfnamefont {M.}~\bibnamefont
  {Jara}},\ }\href {https://doi.org/10.1007/s10955-008-9595-y} {\bibfield
  {journal} {\bibinfo  {journal} {Journal of Statistical Physics}\ }\textbf
  {\bibinfo {volume} {132}},\ \bibinfo {pages} {1135} (\bibinfo {year}
  {2008})}\BibitemShut {NoStop}%
\bibitem [{\citenamefont {Locatelli}\ \emph {et~al.}(2016)\citenamefont
  {Locatelli}, \citenamefont {Pierno}, \citenamefont {Baldovin}, \citenamefont
  {Orlandini}, \citenamefont {Tan},\ and\ \citenamefont
  {Pagliara}}]{Locatelli2016}%
  \BibitemOpen
  \bibfield  {author} {\bibinfo {author} {\bibfnamefont {E.}~\bibnamefont
  {Locatelli}}, \bibinfo {author} {\bibfnamefont {M.}~\bibnamefont {Pierno}},
  \bibinfo {author} {\bibfnamefont {F.}~\bibnamefont {Baldovin}}, \bibinfo
  {author} {\bibfnamefont {E.}~\bibnamefont {Orlandini}}, \bibinfo {author}
  {\bibfnamefont {Y.}~\bibnamefont {Tan}},\ and\ \bibinfo {author}
  {\bibfnamefont {S.}~\bibnamefont {Pagliara}},\ }\href
  {https://doi.org/10.1103/PhysRevLett.117.038001} {\bibfield  {journal}
  {\bibinfo  {journal} {Phys. Rev. Lett.}\ }\textbf {\bibinfo {volume} {117}},\
  \bibinfo {pages} {038001} (\bibinfo {year} {2016})}\BibitemShut {NoStop}%
\bibitem [{\citenamefont {Huang}\ \emph {et~al.}(2018)\citenamefont {Huang},
  \citenamefont {Zhang}, \citenamefont {Lo}, \citenamefont {Lu},\ and\
  \citenamefont {Li}}]{HUANG2018}%
  \BibitemOpen
  \bibfield  {author} {\bibinfo {author} {\bibfnamefont {S.}~\bibnamefont
  {Huang}}, \bibinfo {author} {\bibfnamefont {T.}~\bibnamefont {Zhang}},
  \bibinfo {author} {\bibfnamefont {S.}~\bibnamefont {Lo}}, \bibinfo {author}
  {\bibfnamefont {S.}~\bibnamefont {Lu}},\ and\ \bibinfo {author}
  {\bibfnamefont {C.}~\bibnamefont {Li}},\ }\href
  {https://doi.org/https://doi.org/10.1016/j.physa.2018.06.079} {\bibfield
  {journal} {\bibinfo  {journal} {Physica A: Statistical Mechanics and its
  Applications}\ }\textbf {\bibinfo {volume} {509}},\ \bibinfo {pages} {1023}
  (\bibinfo {year} {2018})}\BibitemShut {NoStop}%
\bibitem [{\citenamefont {Capote}\ \emph {et~al.}(2012)\citenamefont {Capote},
  \citenamefont {Alvear}, \citenamefont {Abreu},\ and\ \citenamefont
  {Cuesta}}]{CAPOTE2012}%
  \BibitemOpen
  \bibfield  {author} {\bibinfo {author} {\bibfnamefont {J.}~\bibnamefont
  {Capote}}, \bibinfo {author} {\bibfnamefont {D.}~\bibnamefont {Alvear}},
  \bibinfo {author} {\bibfnamefont {O.}~\bibnamefont {Abreu}},\ and\ \bibinfo
  {author} {\bibfnamefont {A.}~\bibnamefont {Cuesta}},\ }\href
  {https://doi.org/https://doi.org/10.1016/j.firesaf.2011.12.008} {\bibfield
  {journal} {\bibinfo  {journal} {Fire Safety Journal}\ }\textbf {\bibinfo
  {volume} {49}},\ \bibinfo {pages} {35} (\bibinfo {year} {2012})}\BibitemShut
  {NoStop}%
\bibitem [{\citenamefont {Jeli\ifmmode~\acute{c}\else \'{c}\fi{}}\ \emph
  {et~al.}(2012)\citenamefont {Jeli\ifmmode~\acute{c}\else \'{c}\fi{}},
  \citenamefont {Appert-Rolland}, \citenamefont {Lemercier},\ and\
  \citenamefont {Pettr\'e}}]{Jelic2012}%
  \BibitemOpen
  \bibfield  {author} {\bibinfo {author} {\bibfnamefont {A.}~\bibnamefont
  {Jeli\ifmmode~\acute{c}\else \'{c}\fi{}}}, \bibinfo {author} {\bibfnamefont
  {C.}~\bibnamefont {Appert-Rolland}}, \bibinfo {author} {\bibfnamefont
  {S.}~\bibnamefont {Lemercier}},\ and\ \bibinfo {author} {\bibfnamefont
  {J.}~\bibnamefont {Pettr\'e}},\ }\href
  {https://doi.org/10.1103/PhysRevE.85.036111} {\bibfield  {journal} {\bibinfo
  {journal} {Phys. Rev. E}\ }\textbf {\bibinfo {volume} {85}},\ \bibinfo
  {pages} {036111} (\bibinfo {year} {2012})}\BibitemShut {NoStop}%
\bibitem [{\citenamefont {Kirchner}\ and\ \citenamefont
  {Schadschneider}(2002)}]{KIRCHNER2002260}%
  \BibitemOpen
  \bibfield  {author} {\bibinfo {author} {\bibfnamefont {A.}~\bibnamefont
  {Kirchner}}\ and\ \bibinfo {author} {\bibfnamefont {A.}~\bibnamefont
  {Schadschneider}},\ }\href
  {https://doi.org/https://doi.org/10.1016/S0378-4371(02)00857-9} {\bibfield
  {journal} {\bibinfo  {journal} {Physica A: Statistical Mechanics and its
  Applications}\ }\textbf {\bibinfo {volume} {312}},\ \bibinfo {pages} {260}
  (\bibinfo {year} {2002})}\BibitemShut {NoStop}%
\bibitem [{\citenamefont {Ragoucy}(2017)}]{RAGOUCY2017}%
  \BibitemOpen
  \bibfield  {author} {\bibinfo {author} {\bibfnamefont {E.}~\bibnamefont
  {Ragoucy}},\ }\href {https://doi.org/10.1088/1742-6596/804/1/012037}
  {\bibfield  {journal} {\bibinfo  {journal} {Journal of Physics: Conference
  Series}\ }\textbf {\bibinfo {volume} {804}},\ \bibinfo {pages} {012037}
  (\bibinfo {year} {2017})}\BibitemShut {NoStop}%
\bibitem [{Note1()}]{Note1}%
  \BibitemOpen
  \bibinfo {note} {In this theorem $\gamma $ stands for the proportionality
  coefficient between the number $M$ of steps undergone by a particle and its
  rank $N$, $M=\gamma N$. As the $N$th particle is our leftmost particle, it
  has to hop $L$ times for the lattice to be empty. Therefore $M \equiv L$ and
  $\gamma = 1/\mu $.}\BibitemShut {Stop}%
\bibitem [{Note2()}]{Note2}%
  \BibitemOpen
  \bibinfo {note} {The two first terms of this asymptotic behaviour would be
  the same for 2 particles both initially located on site 1 of the
  lattice.}\BibitemShut {Stop}%
\bibitem [{\citenamefont {Cividini}\ and\ \citenamefont
  {Appert-Rolland}(2017)}]{Cividini_2017}%
  \BibitemOpen
  \bibfield  {author} {\bibinfo {author} {\bibfnamefont {J.}~\bibnamefont
  {Cividini}}\ and\ \bibinfo {author} {\bibfnamefont {C.}~\bibnamefont
  {Appert-Rolland}},\ }\href {https://doi.org/10.1088/1751-8121/aa72d4}
  {\bibfield  {journal} {\bibinfo  {journal} {Journal of Physics A:
  Mathematical and Theoretical}\ }\textbf {\bibinfo {volume} {50}},\ \bibinfo
  {pages} {265002} (\bibinfo {year} {2017})}\BibitemShut {NoStop}%
\bibitem [{\citenamefont {Locatelli}\ \emph {et~al.}(2015)\citenamefont
  {Locatelli}, \citenamefont {Baldovin}, \citenamefont {Orlandini},\ and\
  \citenamefont {Pierno}}]{PhysRevE.91.022109}%
  \BibitemOpen
  \bibfield  {author} {\bibinfo {author} {\bibfnamefont {E.}~\bibnamefont
  {Locatelli}}, \bibinfo {author} {\bibfnamefont {F.}~\bibnamefont {Baldovin}},
  \bibinfo {author} {\bibfnamefont {E.}~\bibnamefont {Orlandini}},\ and\
  \bibinfo {author} {\bibfnamefont {M.}~\bibnamefont {Pierno}},\ }\href
  {https://doi.org/10.1103/PhysRevE.91.022109} {\bibfield  {journal} {\bibinfo
  {journal} {Phys. Rev. E}\ }\textbf {\bibinfo {volume} {91}},\ \bibinfo
  {pages} {022109} (\bibinfo {year} {2015})}\BibitemShut {NoStop}%
\end{thebibliography}%

\end{document}